\documentclass[epj]{svjour}
\usepackage{graphicx}
\usepackage{amssymb,amsbsy,amsmath}
\newcommand{\bs}{\boldsymbol}
\def\be{\begin{equation}}
\def\ee{\end{equation}}
\def\beq{\begin{eqnarray}}
\def\eeq{\end{eqnarray}}
\def\bc{\begin{center}}
\def\ec{\end{center}} 
\newcommand{\nn}{\nonumber\\}
\journalname{Eur. Phys. J. B}
\begin{document}
\title{Alternating-spin  $S=\frac{3}{2}$ and $\sigma=\frac{1}{2}$  Heisenberg chain  with
three-body  exchange interactions}
\titlerunning{Alternating-spin $(\frac{3}{2},\frac{1}{2})$ Heisenberg chain}
\author{N. B. Ivanov%
\inst{1,2}
 \and S. I. Petrova%
\inst{3}
 \and J. Schnack%
\inst{1}
}                     % Do not remove
%
%\offprints{Nedko B. Ivanov}          % Insert a name or remove this line
%
\institute{Department of Physics, Bielefeld University, P.O. box
  100131, D-33501 Bielefeld, Germany 
\and Institute of Solid
  State Physics, Bulgarian Academy of Sciences, Tzarigradsko
  chaussee 72, 1784 Sofia, Bulgaria
\and Department of Engineering Sciences and Mathematics, University of Applied Sciences,
D-33619 Bielefeld, Germany} 
\date{Received: date / Revised version: date}
% The correct dates will be entered by Springer
%
\abstract{
The promotion of collinear classical spin configurations as well as  the enhanced
tendency towards nearest-neighbor clustering of the quantum spins are 
typical  features of the frustrating  isotropic three-body exchange
interactions in  Heisenberg  spin systems. Based on numerical  density-matrix
renormalization group calculations,
we demonstrate that these extra interactions in the Heisenberg chain
constructed from alternating $S=3/2$     
and $\sigma=\frac{1}{2}$ site spins can generate numerous specific quantum spin states, 
including  some partially-polarized ferrimagnetic states as well as
a doubly-degenerate non-magnetic gapped phase. 
In the non-magnetic region of the phase diagram, the model describes a
crossover between the  spin-1 and spin-2 Haldane-type states.
 % end of PACS codes
} %end of abstract
\maketitle
%

%%%%%%%%%%%%%%%%%%%%%%%%%
\section{Introduction}

The biquadratic spin-spin interactions $\left(\bs{S}_i\cdot\bs{S}_j\right)^2$
 and the three-spin exchange couplings  $\left( \bs{S}_i\cdot\bs{S}_j\right)\left(
\bs{S}_i\cdot\bs{S}_k\right)+h.c.$ 
\linebreak
($|\bs{S}_i|>\frac{1}{2}$, $i\neq j, k$;
$j\neq k$)  naturally appear
in the fourth order of the strong-coupling expansion of the 
two-orbital Hubbard model \cite{michaud1}. Since in this case both types of 
couplings are controlled by one and the same model parameter -- which is
about two orders of magnitude weaker than the principal
Heisenberg coupling -- it might  be a  challenge to identify  experimentally accessible
systems  where the effects of higher-order interactions can be definitely isolated.
Unlike the biquadratic exchange couplings \cite{frustration}, 
so far there is no clear evidence for  effects in real
systems related to three-body exchange interactions, although possible three-body
exchange effects in some magnetic molecules \cite{furrer1,furrer2}
and in the spin-$\frac{5}{2}$ Heisenberg chain CsMn$_x$Mg$_{1-x}$Br$_3$
\cite{falk2} have been discussed.

On the theoretical side, only recently some specific features of the three-body 
exchange interaction in Heisenberg spin models in space dimensions D=1
\cite{michaud1,michaud2,ivanov1,ivanov2} and D=2
\cite{michaud3,wang,TRS:PRB12} have been discussed
in the literature. In particular, two of us (N.B.I and J.S) recently  analyzed the full
quantum phase diagram  of the alternating-spin Heisenberg chain \cite{ivanov1}
defined by the Hamiltonian
\beq\label{h}
{\cal H}&=& J_1\sum_{n=1}^L\bs{S}_{2n}\!\cdot\!\left(
\bs{\sigma}_{2n-1}\!+\!\bs{\sigma}_{2n+1}\right)\nn
&+& J_2\sum_{n=1}^L\left[\left(\bs{S}_{2n}\!\cdot\!\bs{\sigma}_{2n-1}\right)
\left( \bs{S}_{2n}\!\cdot\!\bs{\sigma}_{2n+1}\right)\!+\!
h.c.\right], 
\eeq
in the extremely quantum case of on-site spins $S=1$ and  $\sigma=\frac{1}{2}$. 
Here $J_1=\cos \theta$, $J_2=\sin \theta$ ($0\leq \theta <2\pi$),  
and $L$ denotes the number of unit cells containing two different spins ($S>\sigma$).
The model provides  a simple,  but realistic, example of a Heisenberg system with
three-body exchange interactions. For the class of models with $\sigma=\frac{1}{2}$ the biquadratic
terms $\left(\bs{\sigma}_i\cdot\bs{S}_j\right)^2$ reduce to bilinear
Heisenberg spin-spin interactions, so that Eq.~(\ref{h})
represents already the general form of the alternating-spin
Heisenberg chain with higher-order isotropic exchange interactions.

In this article,  we analyze the quantum phase diagram of the above model for
the pair of local spins $S=\frac{3}{2}$ and $\sigma=\frac{1}{2}$. Our motivation 
for this work follows from a previously  established tendency towards
formation of composite spins from the local  $S$ and $\sigma$
spins in the unit  cell--an effect of the three-body exchange interactions
in the region $\frac{\pi}{4}<\theta<\frac{3\pi}{4}$ of the classical phase diagram, which
is characterized by a macroscopic ($2^L$) degeneracy
of the ground state (GS) \cite{ivanov1}. Therefore, one may expect completely different
phase diagrams for systems with integer and half-integer total spin
($S+\sigma$) in the unit cell, especially in the highly degenerate
classical region. Based on density-matrix renormalization group (DMRG)
simulations, in the next Section  we analyze the quantum phase diagram of the
model (\ref{h}) with $S=\frac{3}{2}$ and $\sigma=\frac{1}{2}$ and discuss different  properties of the phases 
appearing  in the interval $0<\theta<\pi$.  The last Section contains a summary of the results.

%%%%%%%%%%%%%%%%%%%%%%%%%%%%%%%%%%%%%%%%%%%%%%%%%%
\section{Quantum phase diagram}
%%%%%%%%%%%%%%%%%%%%%%%%%%%%%%%%%%%%%%%%%%%%%%%%%%

%===================    figure   =================================
\begin{figure}[ht!]
\centering
\includegraphics*[clip,width=75mm]{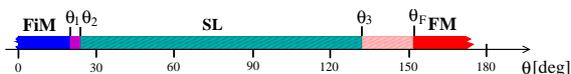}
\caption{(Color online) Quantum phase diagram of the model (\ref{h}) for
$S=\frac{3}{2}$ and $\sigma=\frac{1}{2}$ in the interval $0<\theta <\pi$:
The regions $\theta<\theta_1$ and
$\theta>\theta_F$ are occupied,
respectively, by the N\'{e}el ferrimagnetic (FiM) and ferromagnetic (FM)
phases, whereas the  intervals $\theta_1<\theta<\theta_2$ and 
$\theta_3<\theta<\theta_F$ are occupied by different types of partially-polarized 
magnetic states. A large parameter region,
$\theta_2<\theta<\theta_3$, is occupied by a  non-magnetic
doubly-degenerate gapped phase (SL). The FM point $\theta_F=\pi-\arctan
(\frac{1}{2})\approx 153.43^{\circ}$ is an exact boundary of the FM state,    
$\theta_1=20.1^{\circ}$, $\theta_2=25.5^{\circ}$, and
$\theta_3\approx 132^{\circ}$.
} 
\label{QPD}
\end{figure}
%===================    figure   =================================

\begin{figure}[ht!]
\centering
\includegraphics*[clip,width=70mm]{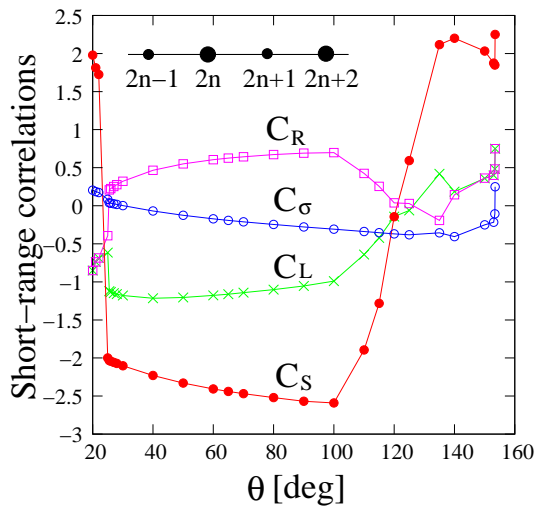}
\caption{(Color online) Short-range spin-spin correlations of the $\left(
\frac{3}{2},\frac{1}{2}\right)$ chain \textit{vs} $\theta$ (DMRG, OBC, L=24). 
$C_L\equiv \langle \bs{\sigma}_{2n-1}\cdot\bs{S}_{2n}\rangle$, 
$C_R\equiv \langle \bs{S}_{2n}\cdot\bs{\sigma}_{2n+1}\rangle$,
$C_{\sigma}\equiv \langle \bs{\sigma}_{2n-1}\cdot\bs{\sigma}_{2n+1}\rangle$, and 
$C_{R}\equiv \langle \bs{S}_{2n}\cdot\bs{S}_{2n+2}\rangle$; $n=1,\ldots,L$.  
} 
\label{SRC}
\end{figure}

The general structure of the phase diagram, as well as the accepted
abbreviations for the phases, are presented in Figure~\ref{QPD}. Most of the
results in this section are obtained through DMRG simulations by performing seven
sweeps and keeping up to 500 states in the last sweep \cite{dmrg1,dmrg2,dmrg3}. 
This ensures a good
convergence  with a discarded weight of the order of
$10^{-8}$ or better. The numerical DMRG analysis of the  lowest 
energy eigenvalues  $E(M)$ in  sectors with a fixed  z component of
the total spin $M$ imply 
(i) a doubly-degenerate non-magnetic gapped GS (SL) 
in the interval $\theta_2<\theta<\theta_3$ and (ii) a number of
specific partially-polarized magnetic states in the intervals
$\theta_1<\theta<\theta_2$  and $\theta_3<\theta<\theta_F$. 
Many features of the phase
diagram in Figure~\ref{QPD} are also encoded in the behavior of the
short-range correlations (SRC) for open boundary conditions (OBC) 
(see  Figure~\ref{SRC}).
%===================    figure   =================================
%===================    figure   =================================
 In particular, most of the phase boundary points in Figure~\ref{QPD} can be
associated with pronounced rearrangements of the SRC. As in the
previously studied extreme quantum case of Eq.~(\ref{h}) with  
$S=1$ and $\sigma=\frac{1}{2}$ \cite{ivanov1}, the basic rearrangements
concern the SRC between the larger $S$ spins, 
whereas -- apart from the region close to the FM point
$\theta_F$ -- the SRC  between 
the $\sigma=\frac{1}{2}$  spins remain almost
constant.\footnote{The equation for the exact FM boundary
  $\theta_F$ for arbitrary spins $S$ and $\sigma$ reads $\cos
  \theta_F+\sigma \left( 2S+1\right)\sin \theta_F=0$
  \cite{ivanov1}.}   
The tendency towards spin clustering is revealed by 
different values of the spin-spin correlators  
$C_L=\langle \bs{\sigma}_{2n-1}\cdot\bs{S}_{2n}\rangle$ 
and 
$C_R=\langle \bs{S}_{2n}\cdot\bs{\sigma}_{2n+1}\rangle$  
in the SL state (see Figure~\ref{SRC}).

%%%%%%%%%%%%%%%%%%%%%%%%%%%%%%%%%%%%%%%%%%%%%%%%%%%%%%%%%%%%%%%%%%%%%%%%%%%%%%%%%
\subsection{\label{sec:} Partially-polarized  magnetic states}
%%%%%%%%%%%%%%%%%%%%%%%%%%%%%%%%%%%%%%%%%%%%%%%%%%%%%%%%%%%%%%%%%%%%%%%%%%%%%%%%%

The established partially-polarized  magnetic states in  the intervals $\theta_1<\theta<\theta_2$ and
$\theta_3<\theta<\theta_F$ do not appear in the  classical phase diagram. 
The critical FiM  phase  in the first interval is identical to the
partially-polarized phase discussed for the extreme quantum case
$(S,\sigma)=(1,\frac{1}{2})$ \cite{ivanov1}: It is characterized by a monotonically 
decreasing magnetization from $m_0=(S-\sigma)=1$ at $\theta=\theta_1$ 
down to $m_0=0$ at the phase boundary $\theta_2$ with the non-magnetic phase. 
At the phase boundary $\theta_1$ the gap of the AFM branch of excitations 
$\Delta_A=E(M_0+1)-E(M_0)$ vanishes and the system becomes critical. Here  
$M_0=(S-\sigma)L$ corresponds to the GS of the Lieb-Mattis  FiM.
Unlike the extreme quantum case, where the phase boundary $\theta_2$ marks
 the transition to a gapless critical  phase, here $\theta_2$ is related
with the vanishing of the triplet gap $\Delta_T$ of the non-magnetic phase SL.
Skipping further  discussions on  this interesting FiM critical state, 
we  only mention that  similar partially-polarized (non-Lieb-Mattis-type)
FiM phases have  been identified and studied  in other spin models, as well 
\cite{ivanov3,shimokawa,furuya}. 

Now, let us turn to the magnetic states  stabilized in the interval 
$\theta_3<\theta<\theta_F$ close to the FM point
$\theta_F$.  The  exact phase boundary  $\theta_F$ coincides with one of
the instability points of the one-magnon  FM excitations and is
characterized by a complete softening of the  dispersion function 
in the whole Brillouin zone. As a result, one observes a strong
reconstruction of the FM state for smaller values of $\theta$.  
As a matter of fact,  for $\theta<\theta_F$ we observe a behavior of the
SRC which is  similar to one  in the extreme quantum system
 (see Figure~4b in Ref.~\cite{ivanov1}).
For this reason,  we shall restrict our discussion mainly to the region 
which is extremely close to the FM point $\theta_F$, as it is natural to expect 
that the formation of specific plateau states depends  on the values of the local 
spins:  According to the general rule, the number of
unit cells in the periodic structure $q$ and the magnetic moment per
unit cell $m_0$ of  the plateau states fulfill the equation  
$q(S+\sigma-m_0)=integer$ \cite{oshikawa}.
%--------------------------------------------------
\begin{figure}
\centering
\includegraphics*[clip,angle=000,width=70mm]{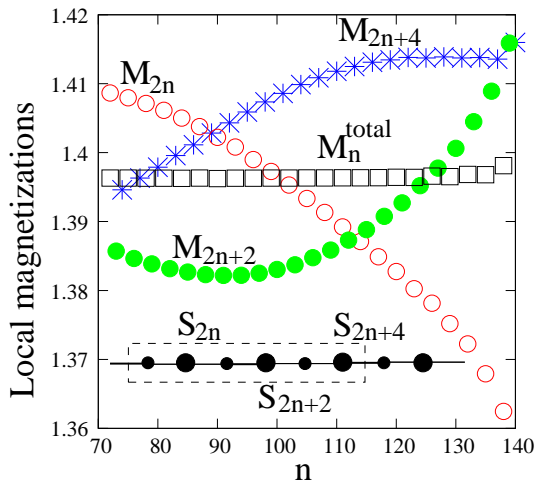}
\caption{(Color online) On-site magnetizations $M_k=\langle S^z_k\rangle$ ($k=2n,2n+2,2n+4$) and
$M_n^{total}\equiv (M_{2n}+M_{2n+2}+M_{2n+4})/3$  as functions of the cell index $n$ 
(DMRG, $\theta=153.4^{\circ}$, $L=144$, OBC). The results  indicate a periodic three-cell
($q=3$) magnetic structure  close to the FM transition point
$\theta_F$. The Inset shows the magnetic supercell containing six
spins (i.e., three unit cells). The total magnetization $M_n^{total}$ in the supercell is constant.
} 
\label{magnetizations}
\end{figure}
%-----------------------------------------

\begin{figure}
\centering
\includegraphics*[clip,width=75mm]{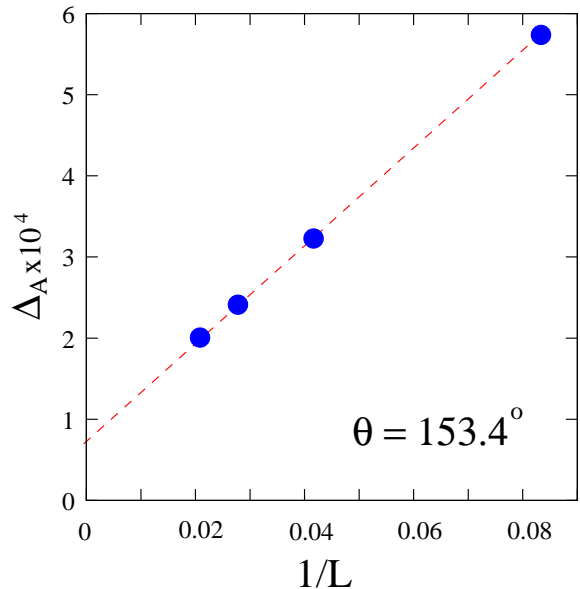}
\caption{
(Color online) Finite-size scaling of  the AFM gap $\Delta_A=E(M_0+1)-E(M_0)$  
above the plateau state with magnetization $m_0\equiv M_0/L=\frac{5}{3}$
(DMRG, OBC). 
} 
\label{gap1}
\end{figure}

In Figure~\ref{magnetizations} we show DMRG results for some local
magnetic moments related to the $S$ spins at $\theta=153.4^{\circ}$, i.e.,  
extremely close to the exact FM boundary
$\theta_F$. The results clearly indicate a periodic magnetic structure with
a period of three unit cells. As required for a plateau state, 
the established magnetization at this
point, $m_0=\frac{5}{3}$, fulfills the mentioned general rule with $q=3$.
The DMRG results for $\Delta_A$ at $\theta=153.4^{\circ}$ shown in
Figure~\ref{gap1} give further support for the suggested plateau state since
the gap is very small but definitely non-zero. Unfortunately, due to strong
finite-size effects, it is difficult to decide if the indicated state is realized
only at $\theta=\theta_F$, or in a small interval close to
this point.     
%===================    figure   =================================
%===================    figure   =================================
Further, as in the extreme quantum case, the 
nearest-neighbor spin-spin correlator 
$C_S$ remains positive and signals a FM ordering of the spin-$S$ subsystem 
in the entire interval $\theta_3<\theta<\theta_F$. The transition 
to a non-magnetic state is accompanied by an abrupt change of the sign 
of the correlator $C_S$. Approaching  the transition point $\theta_3$, 
the boundary effects in open chains become
stronger, so that by using DMRG simulations it is difficult to study the
vicinity of $\theta_3$ and to fix more precisely its position.

%%%%%%%%%%%%%%%%%%%%%%%%%%%%%%%%%%%%%%%%%%%%%%%%%%%%   
\subsection{The non-magnetic SL phase} 
%%%%%%%%%%%%%%%%%%%%%%%%%%%%%%%%%%%%%%%%%%%%%%%%%%%%

The numerical results presented in Figure~\ref{SRC} 
show that for OBC the non-magnetic phase (SL) occupying the
interval $\theta_2<\theta<\theta_3$ is characterized 
by different nearest-neighbor
spin-spin correlations, $C_L\neq C_R$. Excluding some vicinity of 
the phase boundary $\theta_3$, the numerical estimates for $C_L$ 
are located near the eigenvalue $-\frac{5}{4}$ of the operator
$\bs{\sigma}_{2n-1}\cdot\bs{S}_{2n}$. Thus, as a first approximation, 
$\bs{S}_{2n}+\bs{\sigma}_{2n-1}$ ($n=1,\ldots,L$) can be treated as
a spin-1 operator located at the $n-$th unit cell. Respectively, the low-energy sector of the chain
can be analyzed by using the projected  Hamiltonian ${\cal H}_{\text{eff}}=Q^{\dagger}{\cal H}Q$, where
the operator $Q$ is defined as
$$
Q=\prod_{n=1}^LQ_n,\,\,Q_n=\sum_{\alpha_n=0,\pm}|\alpha_n\rangle\langle\alpha_n|\, .
$$
Here $|\alpha_n\rangle$ ($\alpha_n=0,\pm$) are the canonical basis states of the
composite-spin operator $\bs{S}_{2n}+\bs{\sigma}_{2n-1}$ in the spin-1 subspace. In
terms of the Ising states $|S^z_{2n},\sigma^z_{2n-1}\rangle$ the basis
states $|\alpha_n\rangle$ read 
\beq\label{basis}
|0\rangle_n
&=&\frac{1}{\sqrt{2}}\left(\Big{|}-\frac{1}{2},\frac{1}{2}\Big{\rangle}
-\Big{|}\frac{1}{2},-\frac{1}{2}\Big{\rangle}\right)\nonumber \\
|\pm\rangle_n
&=&\frac{1}{2}\left(\mp\sqrt{3}\Big{|}\pm\frac{3}{2},\mp\frac{1}{2}\Big{\rangle}\pm\Big{|}\pm\frac{1}{2},
\pm\frac{1}{2}\Big{\rangle}\right),
\eeq
where for simplicity we have omitted the cell index $n$ on the
right-hand side of the equations. 

Calculating the  matrix elements of the operators $\bs{S}_{2n}$ and
$\bs{\sigma}_{2n-1}$ in the basis (\ref{basis}), 
one obtains  
\be\label{Seff}
Q_n^{\dagger}\bs{S}_{2n}Q_n=\frac{5}{4}\bs{S}_n^{'},\,\,\,
Q_n^{\dagger}\bs{\sigma}_{2n-1}Q_n=-\frac{1}{4}\bs{S}_n^{'},
\ee
where the effective spin-1 operators $\bs{S}^{'}$ are defined as
follows: ${S^{'}}^z=|+\rangle\langle +|-|-\rangle\langle -|$,
${S^{'}}^{+}=\sqrt{2}\left(|+\rangle\langle 0|+|0\rangle\langle
-|\right)$, 
and ${S^{'}}^{-}=\left(
{S^{'}}^{+}\right)^{\dagger}$ for each unit cell.
 Finally, a substitution of Eqs.~(\ref{Seff}) in the expression for ${\cal
H}_{\text{eff}}$ leads to the following effective Hamiltonian
\be\label{heff}
{\cal H}_{\text{eff}}=-\frac{5}{4}J_1L+J_{\text{eff}}
\sum_{n=1}^L\bs{S^{'}}_n\cdot\bs{S^{'}}_{n+1}\, ,
\ee
where $J_{\text{eff}}= \frac{5}{16}\left(\frac{5}{2}J_2-J_1\right)$.

As a matter of fact, Eq.~(\ref{heff}) coincides 
with  the first-order effective Hamiltonian resulting from 
the decoupled-dimer limit defined by the Hamiltonian ${\cal
H}_0=J_1\sum_{n=1}^L\bs{S}_{2n}\!\cdot\!\bs{\sigma}_{2n-1}$. 
Depending on the sign of $J_{\text{eff}}$, the above Hamiltonian
supports a gapped Haldane-type phase ($J_{\text{eff}}>0$) and a partially-polarized
FiM phase ($J_{\text{eff}}<0$). The  transition point at $J_{\text{eff}}=0$ (i.e.,
$\theta=21.8^{\circ}$) corresponds to a completely dimerized GS constructed
from independent spin-1 dimers. This point  is  related to the numerically established 
phase boundary at $\theta=\theta_2$. 
%===================    figure   =================================
\begin{figure}
\centering
\includegraphics*[clip,width=75mm]{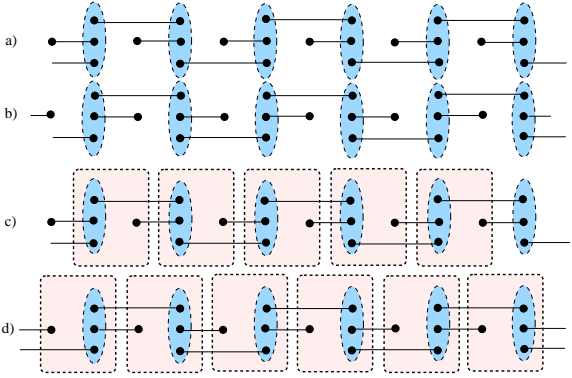}
\caption{(Color online) Valence-bond-solid  picture of the doubly degenerate non-magnetic
phases according to Eq.~(\ref{hmf}) in the limits $|J_F|\ll J_{AF}$ (a,b) and
$|J_F|\gg J_{AF}$ (c,d). The small black dots
denote spin-$\frac{1}{2}$ variables. The lines between two spins $\frac{1}{2}$
denote a singlet bond, whereas  the dashed ellipses and rectangles denote symmetrization of
the spin-$\frac{1}{2}$ variables. The first two (the last two) VBS states approximately
represent ground states of the open spin-1 (spin-2) AFM Heisenberg chain. In the
intermediate region ($|J_F|\approx J_{AF}$) only a part the composite cell spins form spin-2 states.       
} 
\label{VBS}
\end{figure}
%%===================    figure   =================================

In Figure~\ref{VBS}(a,b) we present the  suggested
valence-bond-solid (VBS) states $|\Psi_{L}\rangle$ and
$|\Psi_{R}\rangle$ implementing the discussed 
dimerization features of the GS. Under OBC, there are two such  states depending on the
position of the AFM bond in the three-spin clusters
$\bs{\sigma}_{2n-1}$--$\bs{S}_{2n}$--$\bs{\sigma}_{2n+1}$ ($n=1,\cdots,L$). 
Using the Schwinger representation for  an arbitrary  spin-$S$ operator
with two types of commuting bosons 
(i.e., $S^{+}=a^{+}b$, $S^z=a^{+}a-b^{+}b$, where $a^{+}a+b^{+}b=2S$), 
the related VBS state $|\Psi_{L}\rangle$ for a periodic chain can
be written in the form
$$
|\Psi_{L}\rangle \! =\!
\!\prod_{n=1}^L\!\left(a_{2n}^{+}b_{2n+2}^{+}\! -b\! _{2n}^{+}a_{2n+2}^{+}\right)
\!\!\left(a_{2n-1}^{+}b_{2n}^{+}\! -b\! _{2n-1}^{+}a_{2n}^{+}\right)|0\rangle.
$$
 Here $a_i^{+}a_i+b_i^{+}b_i=2S$ ($2\sigma$) for $i=2n$ ($i=2n-1$) 
 and $|0\rangle$ is the vacuum boson state. Notice that the states $|\Psi_{L}\rangle$ and
 $|\Psi_{R}\rangle$ for an open chain have different number of "dangling" 
spin-$\frac{1}{2}$ free bonds suggesting
 different degeneracy of the GS in the thermodynamic limit. 
This fact may explain the observed automatic selection of one of both states
in the DMRG simulations (see, e.g., Figure~\ref{SRC}) and
considerably complicates the analysis of the low-energy sector
for open chains. 

%%%%%%%%%%%%%%%%%%%%%%%%%%

The dimerization effect of the three-body interaction in the whole interval
$\theta_2<\theta<\theta_3$ can   be approximately studied  by a simple 
decoupling of the three-body  terms in the original Hamiltonian~(\ref{h}):
\beq 
&&\left(\bs{S}_{2n}\!\cdot\!\bs{\sigma}_{2n-1}\right)
\left( \bs{S}_{2n}\!\cdot\!\bs{\sigma}_{2n+1}\right)+\rm{h.c.}\nonumber\\
&=&2C_L \left( \bs{S}_{2n}\!\cdot\!\bs{\sigma}_{2n+1}\right)
+2C_R \left(\bs{S}_{2n}\!\cdot\!\bs{\sigma}_{2n-1}\right)
-2C_LC_R. \nonumber
\eeq
Substituting the above expression in Eq.~(\ref{h}), we obtain  the  following  
"mean-field" spin Hamiltonian with alternating FM-AFM exchange bonds 
\be\label{hmf}
{\cal H}_{\text{MF}}\! =\!\sum_{n=1}^L\left[ J_{AF}\!
\left(\bs{S}_{2n}\!\cdot\!\bs{\sigma}_{2n-1}\right)
\! +\! J_{F}\!
\left(\bs{S}_{2n}\!\cdot\!\bs{\sigma}_{2n+1}\!\right)
\right]\! -\!E_0\, ,
\ee
where $J_{AF}=J_1+2C_RJ_2$, $J_{F}=J_1+2C_LJ_2$, and $E_0=2LJ_2C_LC_R$.
Note that the decoupling procedure violates the translational symmetry 
of the original Hamiltonian (\ref{h}). Since the unit cell in  
Eq.~(\ref{hmf}) is doubled, there is a pair of such  Hamiltonians 
(connected by the symmetry transformation  $J_F\longleftrightarrow J_{AF}$)
related to both types of dimerization functions ($|\Psi_{L,R}\rangle$) introduced above. 
The decoupling procedure can be roughly justified by noting that almost in the whole 
non-magnetic interval the
values of  $C_L$ are  close to the eigenvalue $-\frac{5}{4}$ of the 
operator $\bs{S}_{2n}\cdot\bs{\sigma_{2n-1}}$ (see Figure~\ref{SRC}). This
approximately implies  spin-1 states in the unit cells for each $n=1,\cdots,L$. 
In the  spin-1 subspace, the matrix elements of the thee-body term in Eq.~(\ref{h}) 
coincide with the matrix elements  of  the Heisenberg  term $-\frac{5}{2}J_2\sum_{n=1}^L
\bs{S}_{2n}\cdot\bs{\sigma}_{2n+1}$, so that the basic operator structure of 
Eq.~(\ref{hmf}) can be reproduced. 

\begin{figure}
\centering
\includegraphics*[clip,width=75mm]{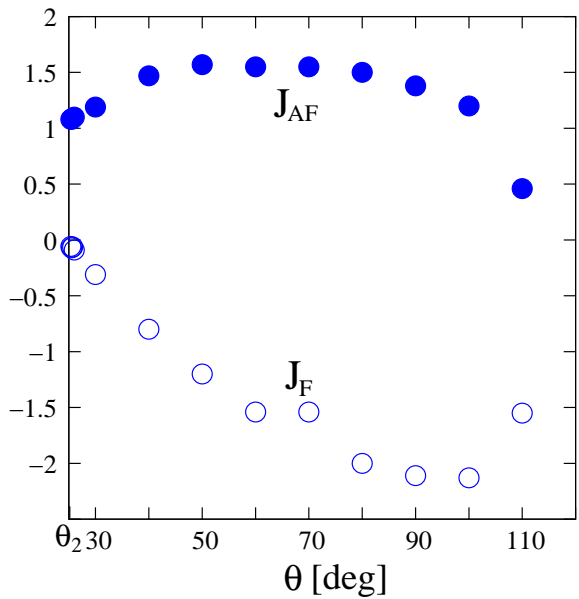}
\caption{(Color online) The effective exchange constants $J_F$ and $J_{AF}$ in
the Hamiltonian ${\cal H}_{\text{MF}}$ as functions of $\theta$.
} 
\label{Jeff}
\end{figure}

In approaching the phase boundary $\theta_2$,
the coupling constant $J_F$ goes to zero (see Figure~\ref{Jeff}), 
so that in this case the decoupled-dimer limit  becomes a valid approximation. 
Up to first order in 
$|J_F|/J_{AF}$, the Hamiltonian ${\cal H}_{\text{MF}}$ is 
equivalent  to the projected spin-1 Hamiltonian~(\ref{heff}) with 
$J_{\text{eff}}=-\frac{5}{16}J_F$ ($J_{\text{eff}}>0$). 
The obtained phase boundary (now defined as $J_F=0$) 
surprisingly well reproduces the numerical estimate
$\theta_2=25.5^{\circ}$.     
%===================    figure   =================================
%%===================    figure   =================================
As far as the parameter  $|J_F|$ increases with $\theta$, it seems relevant
to evaluate the effect of the  second-order perturbation in
$|J_F|/J_{AF}$, as well. However, such a perturbation does not lead to
any qualitative changes of the GS because its  effect is restricted to  a
  small renormalization of  $J_{\text{eff}}$ and to  appearance of an irrelevant 
(FM) next-nearest-neighbor Heisenberg term  in Eq.~(\ref{hmf}). Actually, for 
larger $|J_F|$ it is instructive to analyze  the other decoupled-dimer 
limit of Eq.~(\ref{hmf}) based on non-interacting (FM) spin-2 dimers and
using the small parameter $J_{AF}/|J_{F}|\ll 1$. 
Up to first order in $J_{AF}/|J_{F}|$,  this gives  Eq.~(\ref{heff}), 
but now with the AFM coupling $J_{\text{eff}}=\frac{3}{16}J_{AF}$ and the effective spin-2 operators 
$\bs{S}_n^{'}$. In terms of  VBS states, the formation of local spin-2 states corresponds 
to an additional symmetrization of the cell spins, as shown in
Figure~\ref{VBS}(c,d), without any abrupt change in the topological structure of
the singlet bonds. Therefore, it may be speculated that the transition
 between both dimer limits is realized through a smooth crossover  between 
both Haldane-type gapped states.\footnote{The alternating-bond FM-AFM Heisenberg
chain (\ref{hmf}), 
describing a smooth transition between the spin-1 and spin-2 Haldane phases, 
diserves a special detailed analysis going  beyond the scope of the present
study.}

\begin{figure}
\centering
\includegraphics*[clip,width=75mm]{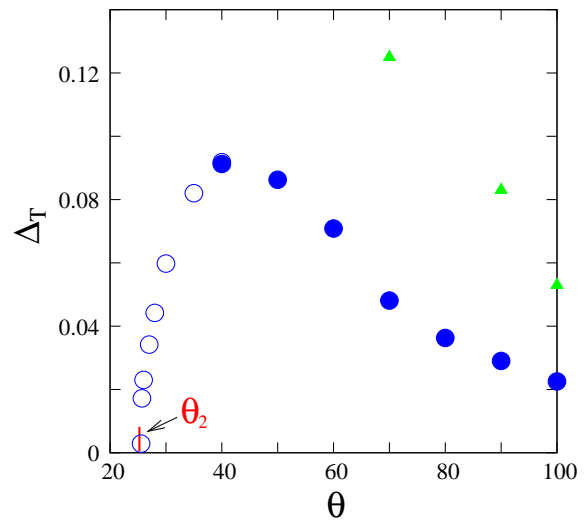}
\caption{(Color online) The  extrapolated (up to $L=60$) thriplet gap $\Delta_T$  
\text{vs} $\theta$ in the SL state calculated
by DMRG under  OBC (open circles) and PBC 
(filled circles). The filled triangles denote the quinted gap under PBC.  
} 
\label{gap}
\end{figure}

In Figure~\ref{gap} we present numerical results for the  triplet energy gap
$\Delta_T$  in the discussed parameter region. The growth of the gap 
approximately up to $\theta\approx 45^{\circ}$ can be related with  the
established increase of the effective exchange constant $J_{\text{eff}}$ in Eq.~(\ref{heff}).   
In accord with the suggested VBS state in Figure~\ref{VBS}(a),  
for OBC  one observes the expected structure of the lowest excited states 
including  a singlet GS, which is  degenerate with the Kennedy edge triplet
in the thermodynamic limit \cite{kennedy}.
The first bulk excitation, related to the Haldane gap, appears as a spin-2 (quintet)
state resulting from the combination of the bulk and Kennedy's
edge triplets.     
On the other hand, for $\theta > 45^{\circ}$ the structure of the lowest excited states
becomes very complicated due to the presence of many parasitic edge
excitations. Namely, as demonstrated in Figure~\ref{VBS}(b,c,d), the number
of free edge spins and their values depend  on (i) the type of 
established VBS states ($|\Psi_{L}\rangle$ or $|\Psi_{R}\rangle$) and (ii)
the increased tendency (with $\theta$) towards  formation of local spin-2
 states. For this reason, for larger $\theta$  the gap  $\Delta_T$ is        
 presented for periodic chains. Since the increase of $|J_f|$
 is restricted  to $|J_f| \lesssim 2$,  the true  spin-2 dimer limit is
not reached. Nevertheless, as far as the  pure spin-2 phase  is characterized by an
extremely small energy gap--$\Delta=0.085(5)J$ according to the DMRG result in Ref.
\cite{uli1}--it is reasonable to admit  that the established
decrease of $\Delta_T$ for $\theta \gtrsim 45^{\circ}$
is connected with  a smooth  crossover  between  the spin-1 and the spin-2 
Haldane-type non-magnetic states.
Finally, due to the extremely small gap $\Delta_T$ and  the large number
of low-lying energy states,  it is difficult to give a precise DMRG estimate 
for the other phase boundary $\theta_3$ and the properties of the GS close to 
this boundary.  
%===================    figure   =================================
%%===================    figure   =================================
%%%%%%%%%%%%%%%%%%%%%%%%%%%%%
\section{Summary}
%%%%%%%%%%%%%%%%%%%%%%%%%%%%%
We have established the general structure of the quantum phase 
diagram of the  alternating-spin $S=\frac{3}{2}$
and $\sigma=\frac{1}{2}$ Heisenberg chain with extra isotropic three-body exchange
interactions. To some extent the established  partially-po\-la\-rized FiM phases
resembler the magnetic phases of the extreme quantum chain 
with alternating spins $S=1$ and $\sigma=\frac{1}{2}$ \cite{ivanov3}, 
apart from the vicinity of the FM point $\theta_F$ where both systems
support different plateau states. On the other hand, due to the clustering
effect of the three-body interactions, both models support completely different quantum phases 
in the non-magnetic region of the phase diagram: the critical phase in the 
$(1,\frac{1}{2})$ model is replaced by  a specific 
doubly-degenerate phase, which can be described as a Haldane-type gapped
state predominantly composed of  effective cell spins with quantum spin numbers $1$ or $2$.
It may be expected that most of the predicted effects and  phases 
persist in higher space dimensions.      

%%%%%%%%%%%%%%%%%%%%%%%%%%%%%%%%%%%%%%%%%%%%%%%%%%%%%%%%%%%%%%%%%%%%%%%%
\section*{Acknowledgment}

This work was supported by the Deutsche Forschungsgemeinschaft
(SCHN 615/20-1) and  by an exchange program between
    Germany and Bulgaria (DAAD PPP Bulgarien 57085392 \&
    DNTS/Germany/01/2). 
S. P. was partially supported by DAAD  (PPP Project ID 57067781), 
the FSP AMMO at the University of  Applied Sciences in  Bielefeld, and
the bulgarian NSF (Grant DFNI I-01/5). 
We thank J.~Ummethum for help with the DMRG program.

%%%%%%%%%%%%%%%%%%%%%%%%%%%%%%%%%%%%%%%%%%%%%%%%%%%%%%%%%%%%%

%%%%%%%%%%%%%%%%%%%%%%%%%%%%%%%%%%%%%%%%%%%%%%%%%%%%%%%%%%%

\begin{thebibliography}{99}
\bibitem{michaud1} 
F. Michaud, F. Vernay, S. R. Manmana,  F. Mila,
Phys. Rev. Lett. \textbf{108}, 127202 (2012)

\bibitem{frustration} 
\textit{Introduction to Frustrated  Magnetism:
Materials, Experiments, Theory}, edited by C. Lacroix, P.
Mendels, F. Mila, Springer Series in Solid-State Sciences, Vol. 164
(2011)  

\bibitem{furrer1} A. Furrer,
Int. J. Mod. Phys. B \textbf{24}, 3653 (2010)

\bibitem{furrer2} 
A. Furrer,  O. Waldmann,
Rev. Mod. Phys. \textbf{85}, 367 (2013)

\bibitem{falk2} 
U. Falk, A. Furrer, J. K. Kjems, H. U. G\"{u}del,
Phys. Rev. Lett. \textbf{52}, 1336 (1984)

\bibitem{michaud2} 
F. Michaud, S. R. Manmana, F. Mila,
Phys. Rev. B \textbf{87}, 140404(R) (2013)

\bibitem{ivanov1}
N. B. Ivanov, J. Ummethum, J. Schnack,
Eur. Phys. J. B \textbf{87}, 226 (2014)

\bibitem{ivanov2}
N. B. Ivanov, J. Schnack,
J. Phys.: Conf. Series \textbf{558}, 012015 (2014)

\bibitem{michaud3}  
F. Michaud,  F. Mila,
Phys. Rev. B \textbf{88}, 094435 (2013)

\bibitem{dmrg1}  
S. R.~White, Phys. Rev. Lett. \textbf{69}, 2863 (1992)

\bibitem{dmrg2}  
J. Ummethum, \emph{Calculation of static and dynamical
  properties of giant magnetic molecules using DMRG},
Ph.D. thesis, Bielefeld University (2012)

\bibitem{dmrg3}  
J. Ummethum, J. Nehrkorn, S. Mukherjee, N. B. Ivanov,
S. Stuiber, Th. Strässle, P. L. W. Tregenna-Piggott, H. Mutka,
G. Christou, O. Waldmann, J. Schnack,
Phys. Rev. B \textbf{86}, 104403 (2012)
 

\bibitem{wang}
Z.-Y. Wang, S. C. Furuya, M. Nakamura, R. Komakura, 
Phys. Rev. B \textbf{88}, 224419 (2013)

\bibitem{TRS:PRB12}
R. Thomale, S. Rachel, P. Schmitteckert, M. Greiter,
Phys. Rev. B \textbf{85}, 195149 (2012)

\bibitem{ivanov3}
N. B. Ivanov, J. Richter,
Phys. Rev. B \textbf{69}, 214420 (2004)

\bibitem{shimokawa}
see, e.g., T. Shimokawa, H. Nakano, J. Kor. Phys. Soc. (SI) \textbf{63}, 591 (2013)
and references therein

\bibitem{furuya}
Sh.S. Furuya, Th. Giamarchi,
Phys. Rev. B \textbf{89}, 205131 (2014) 

\bibitem{oshikawa}
M. Oshikawa, M. Yamanaka,  I. Affleck,
Phys. Rev. Lett. \textbf{78}, 1984 (1997)

\bibitem{kennedy} T. Kennedy, J. Phys. Condens. Matter \textbf{2}, 5737
(1990)

\bibitem{uli1} U. Schollw\"{o}ck, O. Golinelli,  Th. Jolicoeur, 
Phys. Rev. B \textbf{54}, 4038 (1996) 

\end{thebibliography}
\end{document}